\documentclass[onecolumn,prd,showpacs,preprintnumbers,amsmath,amssymb,floatfix]{revtex4}

\usepackage{graphicx}

\usepackage{bm}
\usepackage{amsfonts}
\usepackage{lineno,hyperref}
\usepackage{array}
\usepackage{microtype}
\usepackage{float}
\usepackage{epstopdf}
\usepackage{mathrsfs}
\usepackage{color}
\begin{document}

\title{Rip cosmologies, Wormhole Solutions and Big Trip in the $f(T,\mathcal{T})$ theory of gravity}

\author{Maxime Z. Arouko}
\email{maximearouko55@gmail.com}
\affiliation{D\'epartement de Physique, Universit\'e d'Abomey-Calavi, BP 526 Calavi, Benin}
\author{Ines G. Salako}
\email{inessalako@gmail.com}
\affiliation{Ecole de G\'enie Rural (EGR), 01 BP 55 K\'etou, Benin}
\affiliation{Institut de Math\'ematiques et de Sciences Physiques
(IMSP), 01 BP 613
Porto-Novo, Benin}
\author{A. D. Kanfon}
\email{kanfon@yahoo.fr}
\affiliation{D\'epartement de Physique, Universit\'e d'Abomey-Calavi, BP 526 Calavi, Benin}
\affiliation{Facult\'e des Sciences et Techniques de Natitingou, BP 72, Natitingou, Benin}
\author{ M. J. S. Houndjo}
\email{Sthoundjo@yahoo.fr}
\affiliation{Facult\'e des Sciences et Techniques de Natitingou(FAST), BP 72, Natitingou, Benin}
\affiliation{Institut de Math\'ematiques et de Sciences Physiques
(IMSP), 01 BP 613
Porto-Novo, Benin}
 \author{Etienne Baffou}
\email{baffouh.etienne@yahoo.fr}
\affiliation{Facult\'e des Sciences et Techniques de Natitingou(FAST), BP 72, Natitingou, Benin}
\affiliation{Institut de Math\'ematiques et de Sciences Physiques
(IMSP), 01 BP 613
Porto-Novo, Benin}

\begin{abstract}
Rip cosmological models have been investigated in the framework of $f(T,\mathcal{T})$ theory of gravity, where
$T$ denotes the torsion and $\mathcal{T}$ is the trace of the energy-momentum tensor. These phantom cosmological models 
revealed that at initial epoch a EoS parameter $\omega <-1$ and tends asymptotically  at late phase to $-1$  $(\omega \rightarrow -1)$.
On the other hand, Wormhole Solutions and Big Trip have been subject of an investigation. The wormhole throat radius $R(t)$ and the conditions 
to be satisfied so that produces the Big Trip phenomenon have been discussed.

\end{abstract}

\maketitle
\pretolerance10000
\tableofcontents

\vspace{2pc}

\section{Introduction}\label{int}

Some irrefutable results such as \cite{debamba1,debamba2,debamba3,debamba4,debamba5,debamba6} 
show that the universe our universe is experiencing an accelerated expansion. 
A possible candidate responsible for this current
behavior of the universe is a mysterious energy with negative
pressure which the origin and nature always  stay unelucidated. 
One of the approaches to better understand
this mysterious energy is to modify the gravity. For this purpose,
several modified theories of gravitation such as $f(R)$ gravity, $f(\mathbb{T})$ gravity, 
$f(R,\mathcal{T})$ garvity have been proposed from time to time. 
Similarly, starting from Tele-parallel Theory equivalent of GR
(TEGR)  but not from GTR, one
can consider the matter-coupled modified gravity theory.
One of such theories, namely $ f(T, \mathcal {T}) $ gravity, has been
first proposed by Harko et al\cite{harko/2014}. In this modified theory
the part of the gravitational Lagrangian is taken as an
arbitrary function of torsion scalar $T$ and the trace of the
energy-momentum tensor $\mathcal {T}$. Comparing with the other
theories based on the formalism of curvature or torsion,
$ f(T, \mathcal {T}) $ seems to be a completely different modification
for describing the gravity. \par
Wormholes (WHs) are tunnels or passages that connect two different regions of space-time (or even two distinct universes). 
They have been firstly proposed as a tool for teaching General Relativity (GR) \cite{morris/1988}. Observational evidences for 
such a GR solution have been searched \cite{rahaman/2014}-\cite{toki/2011} and make optimistic the possibility of soon confirming the existence 
of WHs. Moreover, the existence of binary systems containing a WH and a (neutron) star has also been proposed
\cite{dzhunushaliev/2013,dzhunushaliev/2014}.

GR WHs are expected to be filled by exotic matter, that is, matter that does not respect the energy conditions \cite{morris/1988,visser/1996}, 
and may present negative mass (density). An alternative to obtain WH solutions in accordance with the energy conditions is to search for
Morris-Thorne metric \cite{morris/1988} solutions in extended theories of gravity. 

Extended theories of gravity are firstly motivated by the lack of theoretical explanation for some observational/experimental effects 
attained when considering GR as the underlying theory of gravity, such as dark energy \cite{abdalla/2005}, dark matter \cite{akerib/2017,mambrini/2016}, 
missing satellites \cite{klypin/1999,kravtsov/2004}, massive pulsars \cite{antoniadis/2013,demorest/2010}, super-Chandrasekhar white dwarfs 
\cite{howell/2006,silverman/2011}, hierarchy problem \cite{randall/1999,arkani-hamed/1998}, among others. Attempts to solve or evade these shortcomings 
have been developed in different gravitational theories, as one can check Refs.\cite{randall/1999}-\cite{clmaomm/2017}.

Indeed, WHs with non-exotic matter have already been obtained in a multimetric repulsive gravity model \cite{hohmann/2014}, in
higher-order curvature gravity \cite{harko/2013} and in a trace of the energy-momentum tensor squared gravity \cite{ms/2018}, 
for instance. The difficulty in constructing non-exotic matter WHs have led some authors to obtain WHs with arbitrarily small quantities of 
exotic matter \cite{fewster/2005}. WHs particularly satisfying the weak and null energy conditions (WEC and NEC) were obtained, respectively, 
in \cite{mehdizadeh/2015,zangeneh/2015} and \cite{garcia/2011,zangeneh/2014}.

Our intention in the present article is to investigate some Rip cosmological models without any finite time future singularity
in the framework of $f(T,\mathcal{T})$ theory of gravity, where
$T$ denotes the torsion and $\mathcal{T}$ is the trace of the energy-momentum tensor. It will be question in the first section\ref{sec2}, 
a brief review of the $f(T,\mathcal{T})$ theory of gravity. In section\ref{sec3}, we will discuss the physical parameters
where anisotropic and isotropic cases will be addressed. We present in the section\ref{sec4}, the four  different phantom models and 
Wormhole Solutions and Big Trip will be discussed in the section\ref{sec5} where the wormhole throat radius, the Big Trip time and 
the conditions to be satisfied so that the Big Trip phenomenon occurs will presented. Section\ref{sec6} is devoted to a summary and
conclusion.

\section{A brief review of the $f(T,\mathcal{T})$ theory of gravity}\label{sec2}

In GR framework, the metric contains the gravitational potentials responsible for the curvature of space-time. Those potentials can also be represented by the torsion tensor, as in the teleparallel gravity framework (check, for instance, \cite{aldrovandi/2013}). 

The extension of teleparallel gravity is attained when the scalar torsion of teleparallel action is substituted by an arbitrary function of it, namely $f(T)$ \cite{cai/2016}-\cite{cai/2011}. As it is done in teleparallel gravity, the extended versions are also described by the orthonormal tetrads, and their components are defined on the tangent space of each point of the manifold. 

In these theories, the line element is written as

\begin{eqnarray}
ds^2=g_{\mu\nu}dx^\mu dx^\nu=\eta_{ij}\theta^i\theta^j,\label{eq1}
\end{eqnarray}
such that
\begin{eqnarray}
dx^{\mu}=e_{i}^{\;\;\mu}\theta^{i}; \,\quad \theta^{i}=e^{i}_{\;\;\mu}dx^{\mu},\label{eq2}
\end{eqnarray}
with $\eta_{ij}=diag(1,-1,-1,-1)$ being the Minkowskian metric and $\{e^{i}_{\;\mu}\}$ are the components of the tetrad, which satisfy the following identity:
\begin{eqnarray}
e^{\;\;\mu}_{i}e^{i}_{\;\;\nu}=\delta^{\mu}_{\nu},\quad e^{\;\;i}_{\mu}e^{\mu}_{\;\;j}=\delta^{i}_{j}.\label{eq3}
\end{eqnarray}

In GR, one assumes the Levi-Civita's connection, 

\begin{eqnarray}
\mathring{\Gamma }{}_{\;\;\mu \nu }^{\rho } =
\frac{1}{2}g^{\rho \sigma }\left(
\partial _{\nu} g_{\sigma \mu}+\partial _{\mu}g_{\sigma \nu}-\partial _{\sigma}g_{\mu \nu}\right),\label{eq4}
\end{eqnarray}
which preserves the curvature whereas the torsion vanishes. In the teleparallel theory and its extended versions, one keeps the scalar torsion by using the Weitzenb\"{o}ck's connection, defined as:
\begin{eqnarray}
\Gamma^{\lambda}_{\mu\nu}=e^{\;\;\lambda}_{i}\partial_{\mu}e^{i}_{\;\;\nu}=-e^{i}_{\;\;\mu}\partial_\nu e_{i}^{\;\;\lambda}.\label{eq5}
\end{eqnarray}

From the above connection, one obtains the geometric objects of the formalism. The torsion is defined by
\begin{eqnarray}
T^{\lambda}_{\;\;\;\mu\nu}= \Gamma^{\lambda}_{\mu\nu}-\Gamma^{\lambda}_{\nu\mu},\label{eq6}
\end{eqnarray}
from which we define the contorsion as
\begin{eqnarray}
K_{\;\;\mu \nu }^{\lambda} \equiv \Gamma _{\;\mu \nu }^{\lambda }
-\mathring{\Gamma }{}_{\;\mu \nu }^{\lambda}=\frac{1}{2}(T_{\mu }{}^{\lambda}{}_{\nu }
+ T_{\nu}{}^{\lambda }{}_{\mu }-T_{\;\;\mu \nu }^{\lambda}).\label{eq7}
\end{eqnarray}
Then, we can write
\begin{eqnarray}
K^{\mu\nu}_{\;\;\;\;\lambda}=-\frac{1}{2}\left(T^{\mu\nu}_{\;\;\;\lambda}-T^{\nu\mu}_{\;\;\;\;\lambda}+T^{\;\;\;\nu\mu}_{\lambda}\right).\label{eq8}
\end{eqnarray}

The torsion and contorsion tensors are used to define another tensor, as
\begin{eqnarray}
S_{\lambda}^{\;\;\mu\nu}=\frac{1}{2}\left(K^{\mu\nu}_{\;\;\;\;\lambda}+
\delta^{\mu}_{\lambda}T^{\alpha\nu}_{\;\;\;\;\alpha}-\delta^{\nu}_{\lambda}T^{\alpha\mu}_{\;\;\;\;\alpha}\right),\label{eq9}
\end{eqnarray}
such that the torsion scalar can be constructed from torsion and contorsion as 

\begin{eqnarray}
T = S_\sigma^{ \mu\nu} T^\sigma_{ \mu\nu}.\label{eq10}
\end{eqnarray}

In the present article, we will consider an extension of the $f(T)$ theories which also considers terms proportional to the trace of the energy-momentum tensor $\mathcal{T}$ in the action, namely, the $f(T,\mathcal{T})$ theory \cite{harko/2014}. The $f(T,\mathcal{T})$ theory action, for geometrized units, which shall be assumed throughout this work, can be written as

\begin{eqnarray}
\mathbb{S}= \int d^{4}x~~e \left[\frac{T+f(T,\mathcal{T})}{16\pi}+\mathcal{L}_{m} \right],\label{eq11}
\end{eqnarray}
with $\mathcal{L}_m$ being the matter lagrangian.

Varying the action with respect to the tetrad, one obtains the equations of motion \cite{harko/2014}

\begin{eqnarray}\label{eq12}
&&[\partial_\xi(ee^\rho_a
S^{\;\;\sigma\xi}_\rho)-ee^\lambda_a S^{\rho\xi\sigma} T_{\rho\xi\lambda}](1+f_{T})
+ e e^\rho_a(\partial_\xi T)S^{\;\;\sigma\xi}_\rho f_{TT} +\frac{1}{4} e e^\sigma_a T \nonumber \\
&&  =- \frac{1}{4} e e^\sigma_a f(\mathcal{T})  -e e^\rho_a(\partial_\xi \mathcal{T})S^{\;\;\sigma\xi}_\rho f_{T\mathcal{T}}+\frac{f_{\mathcal{T}}}{2}\;(e\,\mathcal{T}^\sigma_{\;\;a}
+ e e^\sigma_a \;p ) + 4\pi\,e\,\mathcal{T}^\sigma_{\;\;a} \;,
\end{eqnarray}
with $f_T = \partial f/\partial T$, $f_{\mathcal{T}} = \partial f/\partial \mathcal{T}$, $ f_{T\mathcal{T}} = \partial^{2}f/\partial T\partial \mathcal{T}$, $f_{TT}  = \partial^{2}f/\partial T^{2}$, $\mathcal{T}^\sigma_{\;\;a}$ is the energy-momentum tensor of the matter field and $p$ its pressure.

By using some transformations, we can establish the following relations:
\begin{eqnarray}\label{eq14}
e^a_\nu e^{-1}\partial_\xi(ee^\rho_a
S^{\;\;\sigma\xi}_\rho)-S^{\rho\xi\sigma}T_{\rho\xi\nu} = -\nabla^\xi S_{\nu\xi}^{\;\;\;\;\sigma}-S^{\xi\rho\sigma}K_{\rho\xi\nu},
\end{eqnarray}

\begin{eqnarray}\label{eq15}
G_{\mu\nu}-\frac{1}{2}\,g_{\mu\nu}\,T
=-\nabla^\rho S_{\nu\rho\mu}-S^{\sigma\rho}_{\;\;\;\;\mu}K_{\rho\sigma\nu},
\end{eqnarray}
with $G_{\mu\nu}$ being the Einstein tensor.

Hence, from the combination of Eqs.(\ref{eq14}) and (\ref{eq15}), the field equations (\ref{eq12}) can be written as

\begin{eqnarray}
G_{\mu\nu}=\kappa_{\mathcal{T}}\;\mathcal{T}_{\mu\nu}^{eff},\label{field:eq1}
\end{eqnarray}
where
\begin{eqnarray}\label{eq20e}
 \kappa_{\mathcal{T}} = \frac{2}{(1+ f_T)},
\end{eqnarray}
and 
\begin{eqnarray}\label{eq20}
\mathcal{T}_{\mu\nu}^{eff} &=&-S^{\rho}_{\;\;\;\mu\nu}\; f_{T\mathcal{T}}\; \partial_{\rho} \mathcal{T}  -
S^{\rho}_{\;\;\;\mu\nu}\;f_{TT}\; \partial_{\rho}T-\frac{1}{4} g_{\mu \nu } f +\frac{1}{4} T\,g_{\mu\nu}\,f_{T}\nonumber\\
&+&\frac{f_{\mathcal{T}}}{2}\;(\mathcal{T}_{\mu \nu}
+ g_{\mu \nu } \;p) + 4\pi \mathcal{T}_{\mu \nu } \;.
\end{eqnarray}
This additive term represents the effective energy-momentum tensor and comes from a minimal coupling with matter. This quantity will vanish 
dès qu'on aurait supprimer la contribution de $\mathcal{T}$ from algebric function $f(T,\mathcal{T})$. 
In other words, this type of coupling (matter-geometry) will generate an additional field that will be perceived
as responsible for the acceleration of the universe. Consequently, we observe the non conservation of the energy impulse tensor.
In order to develop viable cosmological models in according to observational data, an appropriate choice is required.

In the current work, we are focused to study some little rip cosmological models in the extended teleparallel gravity
and for this purpose we user a algebraic function according to observational data \cite{harko/2014}
$f\left(T,\mathcal{T}\right)=\alpha\, T^n\, \mathcal{T} -2  \Lambda$, where $\alpha, n = 0$ and $\Lambda $ 
are arbitrary constants. Thus, the equations (\ref{field:eq1}) and (\ref{eq20e}) yields to

\begin{eqnarray}
 G_{\mu\nu}= \kappa_{\mathcal{T}}\; \mathcal{T}_{\mu\nu}^{eff},\label{feild:3}
\end{eqnarray}
where

\begin{eqnarray}
\mathcal{T}_{\mu\nu}^{eff} =g_{\mu \nu }  \Big( -\frac{\alpha\, \mathcal{T} -2 \Lambda }{4} 
+\frac{\alpha\,p}{2} \Big)
+\;\mathcal{T}_{\mu \nu} \Big(\frac{\alpha\,}{2} + 4\pi \Big)\;.
\end{eqnarray}

For $\alpha=0$, 
the usual  known
in the literature $\Lambda$CDM model is recovered.
We consider the anisotropic metric known as  Locally Rotationally Symmetric Bianchi Type-I model
 \begin{eqnarray}\label{eq:12}
ds^2 = dt^2 - A^2dx^2- B^2(dy^2+dz^2),
\end{eqnarray}
where $A=A(t)$ and $B=B(t)$ are cosmic scale factors.
Note that the flat FRW model is recovered by setting $A(t)=B(t)=a(t)$

In the present investigation, we assume the content of the universe is a cloud of one dimensional 
cosmic strings with string tension density $\xi$ flowing along  $x$-axis. 

Thus, the energy-momentum tensor presents itself
  as following
\begin{eqnarray}\label{eq:13}
\mathcal{T}_{\mu\nu}=(p+\rho)u_{\mu}u_{\nu} - pg_{\mu\nu}-\xi x_{\mu}x_{\nu},
\end{eqnarray}
with
\begin{eqnarray}\label{eq:14}
u^{\mu}u_{\mu}=-x^{\mu}x_{\mu}=1
\end{eqnarray}
and 
\begin{eqnarray}\label{eq:15}
u^{\mu}x_{\mu}=0.
\end{eqnarray}
 Thus, we can consider $\rho$ as being the contribution of particle energy density $\rho_p$ and string 
tension density $\xi$. Note that the contribution of string tension density $\xi$ is vanished when by setting $A(t)=B(t)$.\\

The field equations reads
\begin{eqnarray}
6(k+2)\dot{H}+27H^2 &=& (k+2)^2\;\kappa_{\mathcal{T}} \Big \{ ( -\frac{\alpha\, \mathcal{T} -2 \Lambda }{4} 
 +\frac{\alpha\,p}{2} )
+(-p+\xi)\,(\frac{\alpha\,}{2} + 4\pi )  \Big\}, \label{eq:16}\\
3(k^2+3k+2)\dot{H}+9(k^2+k+1)H^2 &=& (k+2)^2\;\kappa_{\mathcal{T}} \Big \{   \Big( -\frac{\alpha\,\mathcal{T} -2 \Lambda }{4} 
 +\frac{\alpha\,p}{2} \Big)
-p\,\Big(\frac{\alpha\,}{2} + 4\pi \Big)  \Big\},\label{eq:17}\\
9(2k+1)H^2 &=& (k+2)^2 \kappa_{\mathcal{T}} \Big \{   \Big( -\frac{\alpha\, \mathcal{T} -2 \Lambda }{4} 
 +\frac{\alpha\,p}{2} \Big)
+\rho\,\Big(\frac{\alpha\, }{2} + 4\pi \Big)  \Big\}.\label{eq:18}
\end{eqnarray}
where
\begin{eqnarray}\label{torsion}
 T=-2 \left(2\frac{\dot{A} \dot{B}}{AB}+ \frac{\dot{B}^2}{B^2} \right)
\end{eqnarray}

\begin{eqnarray}
 \mathcal{T}= \rho + \xi-2p
\end{eqnarray}

The parameter $k$ provides information about the anisotropic behaviour of the model in the time. Note that by considering
the isotropic model is recovered. In order to investigate on  the isotropization phenomenon, we define the Hubble parameters in the direction of  $[x,y,z] $
\begin{eqnarray}
 H_x = \frac{\dot{A}}{A}\,,\, H_y = \frac{\dot{B}}{B} \,,\,
 H_z = \frac{\dot{B}}{B}           \label{ines1}\,.
\end{eqnarray}
The generalized mean Hubble parameter $H$ is given in the form
\begin{eqnarray}
H& =& \frac{1}{3} \frac{\dot{V}}{V} \cr
  &=&\frac{1}{3}\left( \frac{\dot{A}}{A} +  \frac{\dot{B}}{B} + \frac{\dot{B}}{B}\right) \label{ines2}\,,
\end{eqnarray}
where
\begin{eqnarray}
  V= A \;B^2=a^3  \label{volume}\,,
\end{eqnarray} 
is the spatial volume of the universe and $a$ is the
 scale factor of the universe. The rate of expansion will evaluated by anisotropy parameter given as
\begin{eqnarray}
 \Delta= \frac{1}{3} \sum^{3}_{i=1}\left( \frac{H_i - H}{H} \right)^2\,, \label{ines3}
\end{eqnarray}
where $i =(\;x \;y \;z\;)$. 

Other parameters such that expansion scalar,  
deceleration parameter and Jerk parameter whose
enter online in the processus of isotropization are defined respectively as
\begin{eqnarray}
\text{Expansion scalar:}~~~ \theta &=& u_{;l}^l=\left(\frac{\dot{A}}{A}+2\frac{\dot{B}}{B}\right),\label{eq:19}\\
\text{Deceleration parameter:}~~~ q &=&  -1+\frac{d}{dt}\left(\frac{1}{H}\right),\label{eq:20}\\
\text{Jerk parameter:}~~~  j &=&  \frac{\dddot{a}}{aH^3}=\frac{\ddot{H}}{H^3}-(2+3q).\label{eq:21}
\end{eqnarray}

\section{Physical parameters}\label{sec3}

Current data show us that the Universe is supposed to be homogeneous and isotropic at large scales this is not the case when a local analysis is done.
In such a situation, the anisotropic effects can not be explained by making use of the usual flat FRW model. In the paper, the parameter $k$ 
express this anisotropic universe and we can note that  a flat FRW model is recovered when $k=1$.
\subsection{Anisotropic case}
In indor to explain the phenomenon of isotropization we can  establish respectively some Physical parameters such as pressure, energy density and string tension density from the field 
 equations \eqref{eq:16}-\eqref{eq:18} dependent of   Hubble parameter and  anisotropic parameter
 $k$ 
 \begin{eqnarray}
 P&=& \frac{9 H^2 \left(-16 \left(1+k+k^2\right) \pi +(-3+(-1+k) k) \alpha \right)+(2+k)^2 (16 \pi +\alpha ) \Lambda -3 (2+k)
(16 (1+k) \pi +\alpha -k \alpha ) \dot{H}}{2 (2+k)^2 (4 \pi -\alpha ) (16 \pi +\alpha )}\cr \label{eq:22}
\rho&=& \frac{9 H^2 (16 (1+2 k) \pi +3 (1+(-1+k) k) \alpha )-(2+k)^2 (16 \pi +\alpha ) \Lambda +3 (2+k) (5+3 k) \alpha  \dot{H}}{2
(2+k)^2 (4 \pi -\alpha ) (16 \pi +\alpha )} \cr    \label{eq:23}
\xi&=&-\frac{6 (-1+k) \left(3 H^2+\dot{H}\right)}{(2+k) (16 \pi +\alpha )}   \label{eq:24}
\end{eqnarray}


Consequently, the state parameter (EoS), $\omega=\frac{p}{\rho}$ yields to 
\begin{eqnarray}
\omega=-1 + \frac{12 (4 \pi -\alpha ) \left(-3 H^2 (-1+k) k-(1+k) (2+k) \dot{H}\right)}{9 H^2 (16 (1+2 k) \pi +3 (1+(-1+k) k) \alpha )-(2+k)^2 (16 \pi
+\alpha ) \Lambda +3 (2+k) (5+3 k) \alpha  \dot{H}}   \label{eq:25}
\end{eqnarray}
	
 By considering the previous  expressions of the pressure and energy density we can evaluat the following quantity as 
 \begin{eqnarray}
 \rho+p=-\frac{6 \left(3 H^2 (-1+k) k+(1+k) (2+k) \dot{H}\right)}{(2+k)^2 (16 \pi +\alpha )}  \label{eq:31}
 \end{eqnarray}
 
we can remark for $\alpha \rightarrow 0$, the usual $\Lambda$CDM model is recovered, consequently
Consequently the pressure and EoS parameter becomes $p=-\rho$ and $\omega=-1$. A model that describes an accelerating expanded phantom-like
meets the following requirements as $\dot{H} >0, t>0$ and hence 
the weak energy condition $\rho+p\geq 0; ~ \rho\geq 0$ is not satisfied. In view of equation \eqref{eq:31}, 
the satisfaction of these conditions requires an  appropriate 
choice of parameters $k$ and $\alpha$ in order to preserve the anisotropic nature of the universe.

%

By considering the General Relativity limit case $\alpha \rightarrow 0$, the EoS parameter reads
\begin{eqnarray}\label{eq:32}
\omega=-1+ \frac{-9 H^2 (-1+k) k-3 (1+k) (2+k) \dot{H}}{9 H^2 (1+2 k)-(2+k)^2 \Lambda } .
\end{eqnarray}

In the absence of a cosmological constant it yields
\begin{eqnarray}\label{eq:33}
\omega=-1+ -\frac{3 H^2 (-1+k) k+(1+k) (2+k) \dot{H}}{3 H^2 (1+2 k)} .
\end{eqnarray}


\subsection{Isotropic case}
We develop in this section, the  isotropic model for $k=1$. Therefore,  
the EoS parameter takes the form 
\begin{eqnarray}\label{eq:35}
\omega=-1+ \frac{8 (-4 \pi +\alpha ) \dot{H}}{(16 \pi +\alpha ) \left(3 H^2-\Lambda \right)+8 \alpha  \dot{H}} 
\end{eqnarray}

In particular case i.e $\alpha \rightarrow 0$ and $\Lambda_0 \rightarrow 0$, 
the EoS parameter yields to  the FRW model
\begin{eqnarray}\label{eq:36}
\omega=-1 - \frac{2}{3} \frac{\dot{H}}{H^2} .
\end{eqnarray}

Consequently, the  weak energy condition in this case becomes
\begin{eqnarray}\label{eq:37}
\rho+p= -\frac{4 \dot{H}}{16 \pi +\alpha }.
\end{eqnarray}

This model responds perfectly to a model that describes an accelerating expanded phantom-like
universe $\dot{H}>0$ and $ w < -1$.

\section{Rip cosmologies}\label{sec4}

\subsection{Little Rip}
Little Rip model has been subject a renewed interest and interesting results
have been obtained\cite{Framp2011, Framp2012}. 
We define respectively the  Hubble parameter and  the scale factor as follows 
\begin{eqnarray}\label{eq:39}
H=H_0e^{\lambda t},~~~~~~~~~~ H_0>0,~~\lambda >0 
\end{eqnarray}
\begin{eqnarray}
a = a_0~ exp\left[\frac{H_0}{\lambda}\left(e^{\lambda t}-e^{\lambda t_0}\right)\right].
\end{eqnarray}
where $a_0$ is evaluated scale factor at the current epoch $t_0$. 

By considering (\ref{volume}) and (\ref{torsion}), we can deduce
\begin{eqnarray}
 A &=&  a_0^2 e^{\frac{2 H_0 }{\lambda }\left( e^{\lambda  t} - e^{\lambda  t_0 }\right)}        \\
 B &=& a_0^{\frac{1}{2}} e^{\frac{1}{2}\frac{H_0 }{\lambda }\left(e^{\lambda  t} - e^{\lambda  t_0 }\right)}           \\
 T &=&    -\frac{9}{2} e^{2 t \lambda } H_0^2.         
\end{eqnarray}

Thus, we can see that a inertial force is produced taking into account that the Hubble rate parameter increases exponentially with time.
A particle with mass m at a given point will be subject to an inertial force is given by
Frampton et al. (2012) of the form
\begin{eqnarray}
 F_i &=& m\,l\,(H^2 + \dot{H})\cr
    &=& m\,l\,(H^2_0\,e^{2\lambda t} + H_0\,\lambda \,e^{\lambda t})
\end{eqnarray}
We can remark the principal characteristic of the Little Rip model is the fact that for the time $t \rightarrow \infty$,  
the inertial force $F_i \rightarrow \infty$. From (\ref{eq:39}), we can deduce respectively for the Little Rip model the 
deceleration parameter and the jerk parameter 
\begin{eqnarray}
q &=& -1-\frac{\lambda}{H_0}e^{-\lambda t},\label{eq:40}\\
j &=& 1+\frac{3\lambda}{H_0}e^{-\lambda t}+\left(\frac{\lambda}{H_0}\right)^2e^{-2\lambda t}.\label{eq:41}
\end{eqnarray}
Its remark that this two parameters previous defined tend towards respectively to $-1$ and $1$. The deceleration parameter $q$ evaluated 
at current epoch yields to  
\begin{eqnarray}
 q_0=-1-\frac{\lambda}{H_0}e^{-\lambda t_0},
\end{eqnarray}
 which shows
that $q_0 <-1$. By considering the $\Lambda$CDM model, we remark that the  jerk parameter evaluated 
at current epoch yields to  $j_0=1$. However, by considering the Little Rip model, the jerk parameter evaluated 
at current epoch yields to
\begin{eqnarray}\label{jerk1}
 j_0=1+\frac{3\lambda}{H_0}e^{-\lambda t_0}+\left(\frac{\lambda}{H_0}\right)^2e^{-2\lambda t_0}.
\end{eqnarray}
From (\ref{jerk1}), we can see $j^{RP}_0$ evaluated for Little Rip model greater than that  $j^{\Lambda CDM}_0$ evaluated for $\Lambda$CDM model.   

By considering $\dot{H}=\lambda H >0$, we determine EoS parameter $\omega_{LR}$ for the LR model by replacing  (\eqref{eq:39}) into ( \eqref{eq:25})
\begin{eqnarray}
\omega^{a}_{LR} =-1 -\frac{12 e^{t \lambda } (4 \pi -\alpha ) H_0 \left((1+k) (2+k) \lambda +3 e^{t \lambda } (-1+k) k H_0\right)}{-(2+k)^2 (16 \pi +\alpha
) \Lambda +3 e^{t \lambda } H_0 \left((2+k) (5+3 k) \alpha  \lambda +3 e^{t \lambda } (16 (1+2 k) \pi +3 (1+(-1+k) k) \alpha ) H_0\right)}\label{eq:42}
\end{eqnarray}

From (\ref{eq:42}), its can see that the function $\omega_{LR}$ depends on the anisotropic parameter $k$, 
the coupling constant $\alpha$, the parameters of the scale factors $\lambda$ and $H_0$. We evaluat this function respectively 
at  an initial and epoch ($t\rightarrow 0$) and at a late phase ($t\rightarrow \infty$) as
\begin{eqnarray}
W^{a}_{LR}(t\rightarrow0)=-1 -\frac{12 (4 \pi -\alpha ) H_0 \left((1+k) (2+k) \lambda +3 (-1+k) k H_0\right)}{-(2+k)^2 (16 \pi +\alpha ) \Lambda +3 H_0 \left((2+k)
(5+3 k) \alpha  \lambda +(48 (1+2 k) \pi +9 (1+(-1+k) k) \alpha ) H_0\right)} \label{eq:44}
\end{eqnarray}

\begin{eqnarray}\label{eq:45}
  \omega^{a}_{LR}(t\rightarrow \infty)=-1-\frac{4(4\pi-\alpha)(-1+k)k}{16\pi(1+2k)+3\alpha^{2}(-1+k)k} .
\end{eqnarray}
The analysis EoS parameter for Little Rip model i.e (\ref{eq:44}) and (\ref{eq:45}) at an initial epoch reveal a phantom phase with $\omega_{LR}<-1$ 
However The analysis at late phase shows that  $\omega_{LR} \rightarrow -1$.

Now, we can concern ourselves to isotropic case by looking at the behavior of $\omega_{LR}$ for the little Rip model. Thus, the EoS parameter becomes
\begin{eqnarray}\label{eq:46}
\omega^{i}_{LR}=-1 -\frac{8 e^{t \lambda } (4 \pi -\alpha ) \lambda  H_0}{-(16 \pi +\alpha ) \Lambda +e^{t \lambda } H_0 \left(8 \alpha  \lambda +3 e^{t
\lambda } (16 \pi +\alpha ) H_0\right)}  ,
\end{eqnarray}
which tends  to $-1$ for $t\rightarrow \infty$. By considering  the  General Relativity case i.e $\alpha \rightarrow0$, the EoS parameter becomes
\begin{eqnarray}\label{eq:47}
\omega^{i(GR)}_{LR}=-1 +\frac{2 e^{t \lambda } \lambda  H_0}{\Lambda -3 e^{2 t \lambda } H_0^2} .
\end{eqnarray}
and for 
$\Lambda \simeq 0$, we obtain
\begin{eqnarray}\label{eq:47'}
\omega^{i(GR)}_{LR}=-1-\frac{2 e^{-t \lambda } \lambda }{3 H_0} .
\end{eqnarray}
\subsection{Pseudo Rip}

Pseudo rip known through its phantom behaviour with no finite time singularity is characterized by a hubble parameter defined by\cite{Framp2012}
\begin{eqnarray}\label{eq:48}
H=H_0-H_1e^{-\lambda t},
\end{eqnarray}
where
$H_0, H_1$ and $\lambda$ are positive constants with $H_0 > H_1$. 

We can remark the principal characteristic of the Pseudo Rip model is the fact that for the time $t \rightarrow \infty$,  
a de Sitter universe is recovered i.e $H\rightarrow H_0$. Furthermore, this model presents phantom behaviour i.e 
\begin{eqnarray}\label{der}
 \dot{H}=\lambda H_1e^{-\lambda t}=\lambda(H_0-H) >0
\end{eqnarray}
From (\ref{eq:48}), we can determine the scale factor  as following
\begin{eqnarray}\label{eq:49}
a=a_0~exp\left[H_0(t-t_0)+\frac{H_1}{\lambda}\left(e^{-\lambda t}-e^{-\lambda t_0}\right)\right].
\end{eqnarray}
By make using (\ref{volume}), we obtain
\begin{eqnarray}
 A &=&  a_0^2 \; e^{\frac{2\,H_1 }{\lambda } \left(e^{\lambda  t} - e^{\lambda  t_0 } \right)+ 2\,H_0(t-t_0)}         \\
 B &=& a_0^{\frac{1}{2}} \; e^{\frac{H_1 }{2\,\lambda } \left(e^{\lambda  t} - e^{\lambda  t_0 } \right)+ \frac{1}{2}\,H_0(t-t_0)}            \\
 T &=&    -\frac{9}{2} e^{-2 t \lambda} \Big( -e^{t \lambda } H_0+ H_1 \Big)^2        
\end{eqnarray}
By make using (\ref{eq:48}), the inertial force for Pseudo Rip model is equal to

\begin{eqnarray}
 F_i = m\,l\,\Big(\lambda H_1e^{-\lambda t} + (H_0-H_1e^{-\lambda t})^2 \Big)
\end{eqnarray}
In this case the inertial force is limited, i.e for the time $t \rightarrow \infty$, $F_i \rightarrow m\,l\, H^2_0$.
 From (\ref{eq:48}), we can deduce respectively for the Pseudo Rip model the 
deceleration parameter and the jerk parameter 
\begin{eqnarray}
q &=&-1-\frac{\lambda H_1e^{-\lambda t}}{\left(H_0-H_1e^{-\lambda t}\right)^2},\label{eq:50} \\
j &=& 1-\frac{\lambda H_1e^{-\lambda t}\left[\lambda+3(H_0-H_1e^{-\lambda t})\right]}{\left(H_0-H_1e^{-\lambda t}\right)^3}.\label{eq:51}
\end{eqnarray}

Its remark easily at late epoch that this two parameters previous defined tend towards respectively to $-1$ and $1$ but these parameters evaluated 
at initial epoch yields respectively

\begin{eqnarray}
q(t\rightarrow 0)&=&-1-\frac{\lambda H_1}{\left(H_0-H_1\right)^2} \cr
j(t\rightarrow 0)=1-\frac{\lambda H_1\left[\lambda+3(H_0-H_1)\right]}{\left(H_0-H_1\right)^3}
\end{eqnarray}
 On the other hand, we observe the singularities at 
\begin{eqnarray}
 t=ln\left(\frac{H_1}{H_0}\right)^{\frac{1}{\lambda}}.
\end{eqnarray}
By make using (\ref{der}), we determine EoS parameter $\omega_{PR}$ for the PR model by replacing  (\eqref{eq:48}) into ( \eqref{eq:25})
The EoS parameter for the PR model can be obtained as
\begin{eqnarray}
W^{a}_{PR}(t)=-1+ 
\frac{12 (4 \pi -\alpha ) \left(-e^{-t \lambda } (1+k) (2+k) \lambda  H_1-3 (-1+k) k \left(H_0-e^{-t \lambda } H_1\right){}^2\right)}{-(2+k)^2
(16 \pi +\alpha ) \Lambda +3 e^{-t \lambda } (2+k) (5+3 k) \alpha  \lambda  H_1+9 (16 (1+2 k) \pi +3 (1+(-1+k) k) \alpha ) \left(H_0-e^{-t \lambda
} H_1\right){}^2} 
\label{eq:52}
\end{eqnarray}

Thus, the EoS parameter for the Pseudo Rip model evaluated
at  an initial  epoch ($t\rightarrow 0$) and at a late phase ($t\rightarrow \infty$) yields respectively

\begin{eqnarray}
W^{a}_{PR}(t\rightarrow 0)=-1 -\frac{12 (4 \pi -\alpha ) \left(3 (-1+k) k H_0^2-6 (-1+k) k H_0 H_1+H_1 \left((1+k) (2+k) \lambda +3 (-1+k) k H_1\right)\right)}{-(2+k)^2
(16 \pi +\alpha ) \Lambda +9 (16 (1+2 k) \pi +3 (1+(-1+k) k) \alpha ) \left(H_0-H_1\right){}^2+3 (2+k) (5+3 k) \alpha  \lambda  H_1} \label{eq:53}
\end{eqnarray}

and at a late phase ($t\rightarrow \infty$)
\begin{eqnarray}\label{eq:54}
 \omega^{a}_{PR}(t\rightarrow \infty)=-1-\frac{36(4\pi-\alpha)(-1+k)kH_{0}^{2}}{144(1+2k)\pi H_{0}^{2}+27
[1+(-1+k)k]\alpha H_{0}^{2}-(2+k)^{2}(16\pi-\alpha)} .
\end{eqnarray}

The analysis EoS parameter for Pseudo Rip model i.e (\ref{eq:53}) and (\ref{eq:54}) at an initial epoch reveal a phantom phase with $\omega_{PR}<-1$ 
However The analysis at late phase shows that  $\omega_{PR} \rightarrow -1$.
 we can concern ourselves to isotropic case by looking at the behavior of $\omega_{PR}$ for the Pseudo Rip model. Thus, the EoS parameter becomes

\begin{eqnarray}
W^{i}_{PR}(t)=-1-\frac{72 e^{-t \lambda } (4 \pi -\alpha ) \lambda  H_1}{-9 (16 \pi +\alpha ) \Lambda +72 e^{-t \lambda } \alpha  \lambda  H_1+27 (16
\pi +\alpha ) \left(H_0-e^{-t \lambda } H_1\right){}^2}\label{eq:55}
\end{eqnarray}

which asymptotically tends to $-1$ for $t\rightarrow \infty$. By considering  the  General Relativity case i.e $\alpha \rightarrow0$, 
the EoS parameter reads

\begin{eqnarray}\label{eq:56}
\omega^{i(GR)}_{PR}=-1 -\frac{2 e^{-t \lambda } \lambda  H_1}{-\Lambda +3 \left(H_0-e^{-t \lambda } H_1\right){}^2}.
\end{eqnarray}
and $\Lambda\simeq 0$, we have 
\begin{eqnarray}\label{eq:56}
\omega^{i(GR)}_{PR}=-1 -\frac{2 e^{t \lambda } \lambda  H_1}{3 \left(-e^{t \lambda } H_0+H_1\right){}^2} .
\end{eqnarray}
Easily, we remark that tjis model presents a $\omega$-singularity for $t=t_{\omega}=ln\left(\frac{H_1}{H_0}\right)^{\frac{1}{\lambda}}$ when we consider
General Relativity case. 
\subsection{Emergent Little Rip}
Nous presentons dans cette sous section un autre model proposé par Mukherjee et al.\cite{Mukh2006} presentant un comportement phantomique. 
Ce modèle est caractérisé par un facteur d'échelle présentant une solution émergante défini par 

\begin{eqnarray}\label{eq:57}
a(t)=a_i\left(\nu+e^{\mu t}\right)^{\gamma},
\end{eqnarray}
with $a_i, \mu, \nu$ and $\gamma$ are positive constants.
Thus, we can deduce the Hubble parameter as
\begin{eqnarray}\label{eq:58}
H(t)=\frac{\mu\gamma e^{\mu t}}{\nu+e^{\mu t}}.
\end{eqnarray}
Note that for $t\rightarrow \infty$, the scale factor $a\rightarrow\infty$ and the Hubble parameter $H\rightarrow \mu\gamma$. 
Thus, the de Sitter universe
is recovered. This model also presents phantom behaviour i.e 
\begin{eqnarray}\label{eq:59}
\dot{H}=\frac{\mu\gamma e^{\mu t}}{\nu+e^{\mu t}}\left[\mu-\frac{1}{\gamma}\frac{\mu\gamma 
e^{\mu t}}{\nu+e^{\mu t}}\right]=H\left(\mu-\frac{H}{\gamma}\right) > 0
\end{eqnarray}
pour des valeurs bien précises des positive constants $\mu$ and $\gamma$.

By make using (\ref{volume}), we obtain
\begin{eqnarray}
 A &=&  a_0^2 \, \left(\nu + e^{\mu  t}\right)^{2\gamma }      \\
 B &=& a_0^{\frac{1}{2}}\, \left(\nu  + e^{\mu  t}\right)^{ \frac{\gamma }{2} }          \\
 T &=&    -\frac{9\, e^{2\, t \mu }\, \gamma^2 \mu ^2}{2 \left(e^{t \mu }+\nu \right)^2}         
\end{eqnarray}

By make using (\ref{eq:58}) and (\ref{eq:59}), we can determine the  inertial force for the emergent little rip model as following
\begin{eqnarray}
 F_i = m\,l\,\Bigg( \frac{\mu\gamma e^{\mu t}}{\nu+e^{\mu t}}\left[\mu-\frac{1}{\gamma}\frac{\mu\gamma 
e^{\mu t}}{\nu+e^{\mu t}}\right]+ \frac{\mu^2\,\gamma^2\, e^{2\mu t}}{(\nu+e^{\mu t})^2} \Bigg)
\end{eqnarray}

The deceleration parameter and the jerk parameter for this emergent little rip (ELR) model are obtained as
\begin{eqnarray}
q &=& -1-\frac{\nu}{\gamma},\\
j &=& \left(1-\frac{3}{\gamma}+\frac{2}{\gamma^2}\right)+\frac{\mu}{H}+\frac{\mu(\mu-2/\gamma)}{H^2}.
\end{eqnarray}

Its remark easily at late epoch that he deceleration parameter and the jerk parameter for this emergent little rip  model tend towards 
respectively to $-1$ and $1-\frac{2}{\gamma}+\frac{1}{\gamma^2}\left[2+\frac{(\mu-2/\gamma)}{\mu}\right]$ but these parameters evaluated 
at initial epoch yields respectively

\begin{eqnarray}
q(t\rightarrow 0)&=&-1-\frac{\lambda H_1}{\left(H_0-H_1\right)^2} \cr
j(t\rightarrow 0)&=&1+\frac{\nu-2}{\gamma}+\frac{1}{\gamma^2}\left[2+\frac{(\nu+1)^2(\mu-2/\gamma)}{\mu}\right]
\end{eqnarray}
 
By make using (\ref{der}), we determine EoS parameter $\omega_{ELR}$ for the Emergent Little Rip model by replacing 
\eqref{eq:58} into  \eqref{eq:25}
The EoS parameter for the ELR model can be obtained as

\begin{eqnarray}
W^{a}_{ELR}(t)\textsf{}&=&-1+ \left(12 e^{t \mu } (4 \pi -\alpha ) \gamma  \mu ^2 \left(3 e^{t \mu } (-1+k) k \gamma +(1+k) (2+k) \upsilon \right)
\right) \cr
&/&\left(e^{2
t \mu } \left((2+k)^2 (16 \pi +\alpha ) \Lambda -9 (16 (1+2 k) \pi +3 (1+(-1+k) k) \alpha ) \gamma ^2 \mu ^2\right)+e^{t \mu } (2+k) \left(2 (2+k)
(16 \pi +\alpha ) \Lambda -3 (5+3 k) \alpha  \gamma  \mu ^2\right) \upsilon +(2+k)^2 (16 \pi +\alpha ) \Lambda  \upsilon ^2\right)
\label{eq:62}
\end{eqnarray}

From \ref{eq:62}, we can evaluat  the EoS parameter at a late epoch  as following 

\begin{eqnarray}\label{eq:63}
 \omega_{ELR}(t\rightarrow \infty)=-1+\frac{12\gamma \mu^{2}(4\pi-\alpha)[3(-1+k)k\gamma]}{(2+k)^{2}(16\pi+\alpha)\Lambda-9[16(1+2k)\pi+3(1+(-1+k)k)\alpha]\gamma^{2}\mu^{2}}
\end{eqnarray}

Now, 
from (\ref{eq:62}), we can evaluat  the EoS parameter at a initial epoch  as following

\begin{eqnarray}
W^{aniso}_{ELR}(t\rightarrow0)=-1+ -\frac{12 (4 \pi -\alpha ) \gamma  \mu ^2 (3 (-1+k) k \gamma +(1+k) (2+k) \upsilon )}{(1+\upsilon )^2 \left(-(2+k)^2 (16 \pi +\alpha )
\Lambda +\frac{9 (16 (1+2 k) \pi +3 (1+(-1+k) k) \alpha ) \gamma ^2 \mu ^2}{(1+\upsilon )^2}+\frac{3 (2+k) (5+3 k) \alpha  \gamma  \mu ^2 \upsilon
}{(1+\upsilon )^2}\right)}\label{eq:64}
\end{eqnarray}

By make using of previous expressions, we obtain the EoS parameter for isotropic universe
\begin{eqnarray}
W^{iso}_{ELR}(t)=-1+\frac{8 e^{t \mu } (4 \pi -\alpha ) \gamma  \mu ^2 \upsilon }{e^{2 t \mu } (16 \pi +\alpha ) \left(\Lambda -3 \gamma ^2 \mu ^2\right)+2
e^{t \mu } \left(16 \pi  \Lambda +\alpha  \left(\Lambda -4 \gamma  \mu ^2\right)\right) \upsilon +(16 \pi +\alpha ) \Lambda  \upsilon ^2} \label{eq:65}
\end{eqnarray}

which asymptotically approaches to $-1$ as $t\rightarrow \infty$. 

By considering the limit of General Relativistic
for $\alpha\rightarrow0$, we obtain 
\begin{eqnarray}\label{eq:65}
\omega^{iso(GR)}_{ELR}=-1 + \frac{2 e^{t \mu } \gamma  \mu ^2 \upsilon }{e^{2 t \mu } \left(\Lambda -3 \gamma ^2 \mu ^2\right)+2 e^{t \mu } \Lambda  \upsilon +\Lambda
 \upsilon ^2}
\end{eqnarray}

In absence of cosmological constant $\Lambda\simeq 0$, we obtain
\begin{eqnarray}\label{eq:65}
\omega^{iso(GR)}_{ELR}=-1 -\frac{2 e^{-t \mu } \upsilon }{3 \gamma } .
\end{eqnarray}
\subsection{Bouncing with Little Rip}
Bouncing with Little Rip has been subject an investigation by Myrzakulov and Sebastini \cite{Myrza2014}.
The  scale factor for this model is given by 
\begin{eqnarray}\label{eq:66}
a(t)=a_0e^{(t-t_0)^{2n}},
\end{eqnarray}
where $a_0>0$ is the scale factor evaluated at the today time $t_0$. This model is governed by a constant parameter $n\neq 0$ which shows 
the bouncing behaviour.

Thus, we can deduce the  Hubble parameter for this model as following
\begin{eqnarray}\label{eq:67}
H(t)= 2n(t-t_0)^{2n-1}.
\end{eqnarray}
It is clear that this model exhibits a behavior similar to  little rip at late epoch. Taking the first derivative of we obtain
\begin{eqnarray}
 \dot{H}=2n(2n-1)(t-t_0)^{2n-2}. 
\end{eqnarray}
which yields $\dot{H}>0$ for $n>\frac{1}{2}$. For a exponent $n$ assumes positive integral numbers,  This model present a phantom behaviour.
condition is verified when the exponent $n$ assumes positive integral numbers. Likewise, it note that this model
 present a bouncing at $t=t_0$ when  the bouncing scale factor becomes $a_0$. 
It is easily to remark for positive integral values of $n$ when $t\rightarrow \infty$, we obtain $a\rightarrow\infty$ and $H\rightarrow \infty$ .
By make using (\ref{volume}), we obtain
\begin{eqnarray}
 A &=&   a_0^2 \, e^{2(t-t_0)^{2n}}     \\
 B &=&  a_0^{\frac{1 }{2}}\, \, e^{\frac{1 }{2}(t-t_0)^{2n}}          \\
 T &=&     -18 n^2 (-t_0+t)^{-2+4 n}       
\end{eqnarray}

By make using (\ref{eq:67}), the inertial force for the Bouncing with Little Rip model is equal to 

\begin{eqnarray}
 F_i = m\,l\,\Big( 4n^2\,(t-t_0)^{2(2n-1)}+ 2n(2n-1)(t-t_0)^{2n-2} \Big)
\end{eqnarray}

 From (\ref{eq:67}), we can deduce respectively for the Bouncing with Little Rip model the 
deceleration parameter and the jerk parameter 

\begin{eqnarray}
q &=& -1-\frac{2n-1}{2n(t-t_0)^{2n}},\label{eq:68}\\
j &=& 1+\frac{3(2n-1)}{2n(t-t_0)^{2n}}+\frac{(n-1)(2n-1)}{2n^2(t-t_0)^{4n}}.\label{eq:69}
\end{eqnarray}
The deceleration parameter is a negative quantity for $n > \frac{1}{2}$ and evolves to an asymptotic value of $q=-1$. 
The jerk parameter evolves to $j=1$ at late times.

For the BLR model we can calculate the EoS parameter as
\begin{eqnarray}
&& W^{aniso}_{BLR}(t)=-1 +\cr&& \frac{24 n \left(-(1+k) (2+k) (-1+2 n)-6 (-1+k) k n (t-\text{t0})^{2 n}\right) (t-\text{t0})^{-2+2 n} (4 \pi -\alpha )}{6 (2+k) (5+3 k)
n (-1+2 n) (t-\text{t0})^{-2+2 n} \alpha +36 n^2 (t-\text{t0})^{-2+4 n} (16 (1+2 k) \pi +3 (1+(-1+k) k) \alpha )-(2+k)^2 (16 \pi +\alpha ) \Lambda
}\label{eq:70}
\end{eqnarray}

which asymptotically reduces to
\begin{eqnarray}
  \omega_{BLR}(t\rightarrow \infty)=-1+\frac{24n(4\pi-\alpha)[-6(-1+k)kn]}{36n^{2}[16(1+2k)\pi+3(1+(-1+k)k)\alpha]}  \label{eq:71}
\end{eqnarray}
\begin{eqnarray}
&&W^{a}_{BLR}(t\rightarrow0)=-1+ \cr
&&\frac{24 n \left(-(1+k) (2+k) (-1+2 n)-6 (-1+k) k n (-\text{t0})^{2 n}\right) (-\text{t0})^{2 (-1+n)} (4 \pi -\alpha )}{6 (2+k) (5+3 k)
n (-1+2 n) (-\text{t0})^{2 (-1+n)} \alpha +36 n^2 (-\text{t0})^{-2+4 n} (16 (1+2 k) \pi +3 (1+(-1+k) k) \alpha )-(2+k)^2 (16 \pi +\alpha ) \Lambda
} \label{eq:72}
\end{eqnarray}

The EoS parameter for this BLR model in an isotropic universe can be expressed as,
\begin{eqnarray}
W^{i}_{BLR}(t)=-1-\frac{16 n (-1+2 n) (t-\text{t0})^{2 n} (4 \pi -\alpha )}{16 n (-1+2 n) (t-\text{t0})^{2 n} \alpha +12 n^2 (t-\text{t0})^{4 n} (16 \pi
+\alpha )-(t-\text{t0})^2 (16 \pi +\alpha ) \Lambda } \label{eq:73}
\end{eqnarray}

which asymptotically approaches to $-1$ as $t\rightarrow \infty$. 
In the limit of GR with $\alpha \rightarrow0$ 
\begin{eqnarray}\label{eq:72}
\omega^{iso(GR)}_{BLR}=-1  -\frac{4 n (-1+2 n) (t-\text{t0})^{2 n}}{12 n^2 (t-\text{t0})^{4 n}-(t-\text{t0})^2 \Lambda }.
\end{eqnarray}
and $\Lambda\simeq 0$, we have 
\begin{eqnarray}\label{eq:72'}
\omega^{iso(GR)}_{BLR}=-1+ \frac{(1-2 n) (t-\text{t0})^{-2 n}}{3 n}.
\end{eqnarray}
\section{Wormhole Solutions and Big Trip}\label{sec5}
The phenomenon called Big Trip is often observed
  when the size of the wormhole throat  takes volume. This is due to the fact
a strong phantom energy buildup
on the wormhole which can lead to an absorption of the whole universe
before the appearance of the rip phenomenon. 
Now, we will be interested in the determination of the wormhole throat radius and his behaviour under
the phantom energy accumulation.
This wormhole throat radius is determined via the following differential equation for the isotropic case with vanishing cosmological 
constant\citep{Asta2012, Babichev2004}

\begin{eqnarray}\label{eq:73}
\dot{R}= -C_0 R^2(\rho+p).
\end{eqnarray}
when $C_0$ is a positive dimensionless constant.\\
\subsection{I-Little Rip case}
We determine the wormhole throat radius $R(t)$ by combining (\eqref{eq:37}) and (\eqref{eq:39})  into \eqref{eq:73}
\begin{eqnarray}
\frac{1}{R_{LR}(t)}= C_1-\frac{4 C_0 e^{t \lambda } H_0}{16 \pi +\alpha },\nonumber
\end{eqnarray}
where $C_1$ is an integration constant which is determined at big trip time as 
\begin{eqnarray}
 C_1=\frac{2C_0}{16 \pi +\alpha }H_0e^{\lambda t_{Big}}.
\end{eqnarray}
 Thus, we can easily   the wormhole throat radius 
\begin{eqnarray}\label{eq:75}
R_{LR}(t)=\frac{16 \pi +\alpha }{2C_0H_0}\left[e^{\lambda t_{Big}}-e^{\lambda t}\right]^{-1}.
\end{eqnarray}
We can also determine the Big Trip assuming that $R(t_0)=R_0$ through

\begin{eqnarray}\label{eq:76}
t_{Big}=ln\left[e^{\lambda t_0}+\frac{16 \pi +\alpha}{2C_0H_0R_0}\right]^{\frac{1}{\lambda}},
\end{eqnarray}
which leads asymptotically to in General relativity $\alpha\rightarrow 0$
\begin{eqnarray}\label{eq:77}
t^{GR}_{Big} =ln\left[e^{\lambda t_0}+\frac{8\pi}{C_0 H_0R_0}\right]^{\frac{1}{\lambda}}.
\end{eqnarray}
The difference between (\eqref{eq:76}) and (\eqref{eq:77}) reveal  the importance of the coefficient $\alpha$ but especially 
the contribution of the modified gravity theory in the process of Big Trip phenomenon. 

For this Little Rip case, it is necessary to remark that the Big Trip phenomenon must realize when the following condition is verified $t=t_0$ 
\begin{eqnarray}
R_0 >\frac{16 \pi +\alpha}{2C_0H_0}e^{\lambda t_0}.
\end{eqnarray}
\subsection{II-Pseudo Rip case}
By analogy to the same previous method, we determine the wormhole throat radius  for the Pseudo Rip case 
by combining (\eqref{eq:37}) and (\eqref{eq:48})  into \eqref{eq:73}
\begin{eqnarray}
R_{PR}(t)=\frac{16 \pi +\alpha}{2C_0H_1}\left[e^{-\lambda t}-e^{-\lambda t_B}\right]^{-1}.
\end{eqnarray}
Thus, we can deduce  the Big Trip time  for the Pseudo Rip case as
\begin{eqnarray}\label{dd}
t_{Big}=ln\left[e^{-\lambda t_0}-\frac{16 \pi +\alpha}{2C_0H_1R_0}\right]^{-\frac{1}{\lambda}}.
\end{eqnarray}
 As previously, the following condition will must verified at $t=t_0$ for the realization of the Big Trip phenomenon 
\begin{eqnarray}
R_0 >\frac{16 \pi +\alpha}{2C_0H_1}e^{\lambda t_0}.
\end{eqnarray}
In the limit of General Relativity,  the Big Trip time becomes
\begin{eqnarray}\label{dd1}
t^{GR}_{Big}=ln\left[e^{-\lambda t_0}-\frac{8 \pi}{C_0H_1R_0}\right]^{-\frac{1}{\lambda}},
\end{eqnarray}
It note the effect of the coupling constant $\alpha$ through the equations (\eqref{dd}) and (\eqref{dd1}).
\subsection{III-Emergent Little Rip}
 We determine the wormhole throat radius  for the Emergent Little Rip case 
by combining (\eqref{eq:37}) and (\eqref{eq:58})  into \eqref{eq:73} as following
\begin{eqnarray}
R_{ELR}(t)=\frac{16 \pi +\alpha}{2C_0\mu\nu\gamma}\left[\frac{1}{\nu+e^{\mu t}}-\frac{1}{\nu+e^{\mu t_B}}\right]^{-1}.
\end{eqnarray}

Thus, we can determine  the Big Trip time  for the Emergent Little Rip case 
\begin{eqnarray}
t_{Big}=ln\left[\left(\frac{1}{\nu+e^{\mu t_0}}-\frac{16 \pi +\alpha}{2C_0\mu\nu\gamma R_0}\right)^{-1}-\nu\right]^{\frac{1}{\mu}},
\end{eqnarray}
which we deduce at the limit of General Relativity, the Big Trip time 

\begin{eqnarray}
t^{GR}_{Big} =ln\left[\left(\frac{1}{\nu+e^{\mu t_0}}-\frac{8 \pi }{C_0\mu\nu\gamma R_0}\right)^{-1}-\nu\right]^{\frac{1}{\mu}},
\end{eqnarray}
and the condition to be satisfied so that produces the Big Trip phenomenon

\begin{eqnarray}
R_0 > \frac{(16 \pi +\alpha)(\nu+e^{\mu t_0})}{2C_0\mu\nu\gamma}
\end{eqnarray}
and 
\begin{eqnarray}
R_0<\frac{\nu(16 \pi +\alpha)}{2C_0\mu\nu\gamma\left[\nu-(\nu+e^{\mu t_0})\right]}.
\end{eqnarray}
\subsection{IV-Bouncing with Little Rip case}
By combining (\eqref{eq:37}) and (\eqref{eq:67})  into \eqref{eq:73}, We determine respectively the wormhole throat radius  and 
the Big Trip time for the Bouncing with Little Rip case 
 as following

\begin{eqnarray}
R_{BLR}(t)=\frac{16 \pi +\alpha}{4C_0 n}\left[\left(t_B-t_0\right)^{2n-1}-\left(t-t_0\right)^{2n-1}\right]^{-1},
\end{eqnarray}
and 
\begin{eqnarray}
t_B=t_0+\left[\left(t_1-t_0\right)^{2n-1}+\frac{16 \pi +\alpha}{4C_0 n R_1 }\right]^{\frac{1}{2n-1}}.
\end{eqnarray}
We deduce at the limit of General Relativity, the Big Trip time 
\begin{eqnarray}
t_B=t_0+\left[\left(t_1-t_0\right)^{2n-1}+\frac{16 \pi }{4C_0 n R_1}\right]^{\frac{1}{2n-1}}.
\end{eqnarray}
where $R(t_1)=R_1 $
\section{Conclusion}\label{sec6}
In this paper, some phantom models without any Big Rip singularity 
at finite time have been subject of an investigation in the context of $f(T,\mathcal{T})$ theory of gravity, where
$T$ denotes the torsion and $\mathcal{T}$ is the trace of the energy-momentum tensor. These phantom cosmological models 
revealed that at initial epoch a EoS parameter $\omega <-1$ and tends asymptotically  at late phase to $-1$  $(\omega \rightarrow -1)$. 
These models are seems to that of Little Rip models where it remark that for the time $t \rightarrow \infty$,  
a de Sitter universe is recovered i.e $H\rightarrow H_0$. 
Four different phantom models have been investigated where we focused on anisotropic and isotropic universe.
We found that the coupling constant change dynamically the behaviour of EoS parameter. \par
In addition to his studies, Some wormhole solutions have been obtained for these Four different phantom models. 
We have also determined for these phantom models, the wormhole throat radius, the Big Trip time and we have discussed 
we discussed about the conditions to be satisfied so that the Big Trip phenomenon occurs. It should be noted, as previously that 
the coupling constant of $f(T,\mathcal{T})$ theory of gravity affect the Big Trip time when passing the boundary of the General Relativity.

\end{document}